\documentclass[twocolumn,showpacs]{revtex4} \usepackage{graphicx}
\usepackage{dcolumn} \usepackage{bm} \usepackage{amsmath}
\usepackage{amsfonts} \usepackage{amssymb}
\begin{document}
\title{An exactly solvable model of the BCS-BEC crossover}

\author{J.N.\ Fuchs, A. Recati, W. Zwerger}

\affiliation{Institute for Theoretical Physics, Universit\"at Innsbruck, Technikerstrasse 25, A-6020 Innsbruck, Austria}

\date{\today}

\begin{abstract} 

We discuss an integrable model of interacting Fermions in one
dimension, that allows an exact description of the crossover from a
BCS- to a Bose-like superfluid. This model bridges the Gaudin-Yang
model of attractive spin $1/2$ Fermions to the Lieb-Liniger model of
repulsive Bosons.  Using a geometric resonance in the one-dimensional
scattering length, the inverse coupling constant varies from $-\infty$
to $+\infty$ while the system evolves from a BCS-like state through a
Tonks gas to a weakly interacting Bose gas of dimers. We study the
ground state energy, the elementary density and spin excitations, and
the correlation functions.  An experimental realization with cold
atoms of such a one-dimensional BCS-BEC crossover is proposed.
\end{abstract}


\maketitle

Starting with the historic controversy between Bardeen and Schafroth
about the proper explanation of superconductivity \cite{GF}, the
crossover from a Bardeen-Cooper-Schrieffer (BCS) superfluid with
Cooper pairs, whose size is much larger than the inter-particle
spacing, to a Bose-Einstein Condensate (BEC) of molecules composed of
tightly bound Fermion pairs has been a basic issue in many-body
physics \cite{ELND}.  Very recently, with the observation of molecular
condensates near a Feshbach resonance in cold atomic Fermi gases
\cite{GGK}, this problem has regained a lot of attention, since for
the first time the crossover can be studied in detail. While there is
general agreement on the fact that the evolution between the BCS and
Bose-limit is continuous, all existing theories of the crossover are
approximate \cite{Randeria}.  In particular, there is no reliable
description of the most interesting regime near the Feshbach
resonance, where the scattering length diverges. It is therefore of
considerable interest to have an analytically soluble model of the
BCS-BEC crossover. This is what we will provide in this letter for the
particular case of one dimension (1D).

Cold gases allow to realize the BCS-BEC crossover by driving a mixture
of two spin states with an attractive interaction through a Feshbach
resonance, beyond which a bound state appears in the two particle
problem in free space. On the BCS side of the crossover, pairs only
exist in the many-body system due to the Pauli blocking of states
below the Fermi energy. In one and also in two dimensions, the
situation is quite different, however, because any purely attractive
interaction produces a bound state already at the two particle
level. In fact, contrary to the 3D case, its existence is both a
necessary and sufficient condition for a BCS instability
\cite{Randeria2}. As will be shown below, an analog of the 3D
crossover can be achieved in 1D by exploiting a confinement induced
resonance (CIR) in a tight trap where the effective 1D scattering
length exhibits a resonance caused by the mixing with a closed channel
bound state in the trap \cite{Olshanii}.

We start by considering the Hamiltonian of the Gaudin-Yang (GY) model
\cite{GY} of a spin $1/2$ Fermi gas interacting via a short range
potential $g_1\delta(x)$:
\begin{equation}
H=-\frac{\hbar^2}{2m}\sum_{i=1}^{N}\frac{\partial^2}{\partial x_i^2} +
g_1 \sum_{i<j}\delta(x_i-x_j)
\label{Ham}
\end{equation}
where $N$ is the total number of Fermions and $m$ their mass.  The
single dimensionless coupling constant is $\gamma\equiv mg_1 / \hbar^2
n$, where $n\equiv N/L$ is the 1D density.  For attractive
interactions, the Hamiltonian (\ref{Ham}) describes a Luther-Emery
liquid \cite{LE}. When $\gamma \rightarrow 0^{-}$, its ground state is
a BCS-like state with Cooper pairs, whose size is much larger than the
average inter-particle spacing \cite{KO}. The strong coupling regime
with tightly bound molecules is now simply reached by increasing the
magnitude of $\gamma$. The resulting Fermion pairs behave like a hard
core Bose gas, or equivalently like 1D {\it non}-interacting Fermions
\cite{Girardeau}. In this manner, obviously, one never reaches a
weakly interacting BEC as one of the limits of the standard BCS-BEC
crossover in 3D. This is a consequence of the trivial fact that the
two-body potential $g_1 \delta(x)$ has a bound state only when
$g_1<0$, but none when $g_1>0$.  As a result, in the regime $g_1>0$,
the ground state of the Hamiltonian (\ref{Ham}) is that of repulsive
Fermions and thus has nothing in common with a weakly interacting Bose
gas of molecules.

In order to obtain a genuine BCS-BEC crossover in 1D, it is necessary
to exploit a geometric resonance in the 3D two-body scattering problem
with a strong transverse confinement.  For simplicity, we assume the
Fermions to be confined in a waveguide with radial frequency
$\omega_{\perp}/2\pi$ and oscillator length $a_{\perp}\equiv
\sqrt{\hbar/m\omega_{\perp}}$.  Moreover, we restrict ourselves to a
truly 1D situation, where only the lowest transverse mode is occupied,
requiring $\hbar \omega_{\perp}$ to be much larger than the Fermi
energy $\epsilon_F$.  As shown by Bergeman et al. \cite{BMO}, in such
a situation there is always precisely one two-body bound state for the
longitudinal motion with energy $\tilde{\epsilon}_b$, \emph{whatever}
the 3D scattering length $a$.  Apart from this bound state, all the
scattering properties are perfectly described by an effective 1D delta
potential $g_1\delta(x)$ with strength \cite{Olshanii}:
\begin{equation}
g_1=2 \hbar \omega_{\perp}a \left( 1-Aa/a_{\perp}\right)^{-1}
\label{g1olsh}
\end{equation}
As naively expected, an attractive 3D interaction $a<0$ implies a
negative value of $g_1$, associated with a bound state whose energy
$\epsilon_b=-mg_1^2/4\hbar^2$ coincides with the exact bound state
energy $\tilde{\epsilon}_b$ in the limit $a/a_{\perp}\to
0$. Remarkably, $g_1$ and thus the binding energy $\epsilon_b$ remain
finite at a Feshbach resonance ($a=\pm \infty$).  Entering the
positive side $a>0$, however, the vanishing of the denominator at
$a_{\perp}/a=A\simeq 1.0326$ \cite{sqrt2} leads to a CIR, where $g_1$
jumps from $-\infty$ to $+\infty$ just as in a standard 3D Feshbach
resonance, as discussed recently also in \cite{Astra}.  Now, for
$g_1>0$, the short range potential $g_1\delta(x)$ no longer has a
bound state, however it is still there in the 3D problem. Due to the
condition $\hbar \omega_{\perp}\gg \epsilon_F$ (equivalently
$(na_{\perp})^2\ll 1$), the true bound state energy at CIR,
$\tilde{\epsilon}_b=-2\hbar\omega_{\perp}$ \cite{BMO}, is much lower in
energy than any other relevant scales and may formally be taken to
minus infinity.

The 1D analog of a Feshbach resonance driven BCS-BEC crossover in 3D
is therefore described by a modified Gaudin-Yang model (\ref{Ham}),
where for positive $g_1$ the repulsive short range potential is
supplemented by an additional bound state with energy
$\epsilon_b=-\infty$.  After crossing the CIR at $1/\gamma=0$, the
unbreakable Fermion pairs are described by a Lieb-Liniger (LL)
model \cite{LL} of repulsive Bosons. On a formal level, the continuous
evolution from an attractive Fermi to a repulsive Bose gas in one
dimension is implicit in the Bethe Ansatz equations of the
GY and LL models. The ground state for
both $\gamma<0$ and $\gamma >0$ may be obtained from the solution of
the equations:
\begin{eqnarray}
\pi\rho(k) &=& 1+\int_{-K}^{K}\frac{dq}{n}\frac{\gamma
\rho(q)}{\gamma^2+[(k-q)/n]^2} \nonumber \\
\frac{E_0}{N}&=&\frac{\epsilon_b}{2}+2\int_{-K}^{K}\frac{dk}{n}
\rho(k)\frac{\hbar^2k^2}{2m}
\label{BAE}
\end{eqnarray}
where the quasi-momenta distribution $\rho(k)$ is normalized as $\int
dk\rho(k)=n/2$ and $E_0$ is the ground state energy.  For $\gamma<0$,
they reduce to the GY equations of an attractive Fermi gas
while for $\gamma>0$, they reduce to the LL equations of a
gas of dimers \cite{Gaudin}.  This equivalence shows that in the whole
regime $\gamma>0$, the density of Bosons is $n_B=n/2$, their mass is
$m_B=2m$, their coupling $g_B$ is identified with the
one of the LL model and $\gamma\equiv mg_B/\hbar^2n$.  Since
one is now dealing with dimers, the relation between the coupling
constant $g_B$ and the experimentally accessible parameters $a$ and
$a_{\perp}$ is no longer given by (\ref{g1olsh}), however. It
requires a solution of the dimer-dimer scattering problem in the
presence of a transverse confinement. In the limit $a\ll a_{\perp}$,
one can use the free space result for the effective scattering length 
between point like dimers $a_{3B}\approx 0.6\, a$ \cite{Petrov},
leading to $g_B\approx 1.2\hbar\omega_{\perp}\, a$. Close to resonance 
($a_{\perp}/a\gtrsim A$), a plausible \emph{ansatz} is:
\begin{equation}
g_B=\sqrt{2}\hbar \omega_{\perp}a \left( 1-Aa/a_{\perp}\right)^{-1}
\label{gB}
\end{equation}
It follows by assuming that an equation of the form (\ref{g1olsh})
holds also for dimers and the requirement that the resonances match,
while $a_{\perp}$ is reduced by a factor $1/\sqrt{2}$. Note that
(\ref{gB}) corresponds to an effective dimer scattering length
$a_{3B}=a/\sqrt{2}$.

The qualitative physics of the modified Gaudin-Yang model is now
simple to understand: when $1/\gamma \rightarrow -\infty$ (BCS limit),
the system consists of weakly bound Cooper pairs.  The associated
excitation gap is related to the spin sector, i.e. there is a finite
gap between the singlet ground state and the first triplet excited
state. In addition, there are gapless density fluctuations describing
the Bogoliubov-Anderson mode of a neutral superfluid.  At resonance,
when $1/\gamma=0$, the system is a Tonks gas \cite{Girardeau} of
tightly bound dimers \cite{Astra}.  It still exhibits sound modes with
a linear spectrum, however the spin sector has disappeared because the
spin gap diverges.  On the positive side of the resonance, the system
is an interacting Bose gas of tightly bound molecules. Its excitations
are the standard Bogoliubov sound modes, whose velocity vanishes
asymptotically in the weak coupling limit $1/\gamma \rightarrow
+\infty$ (BEC limit).

\begin{figure}[ptb]
\begin{center}
\includegraphics[height=4.5cm] {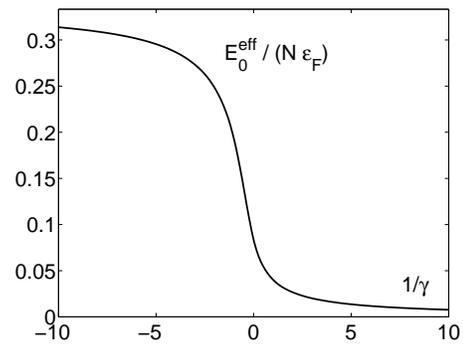}
\end{center}
\caption{Ground state effective energy per particle 
$E_0^{eff}/N=E_0/N-\epsilon_b/2$ [in units of the Fermi energy for the
non-interacting gas $\epsilon_F=\pi^2\hbar^2n^2/8m$] as a function of
$1/\gamma$.}
\label{energy}
\end{figure}

For quantitative results, we start with the ground state where the
number of up and down spins is identical. Subtracting the bound state
contribution, the relevant finite quantity to consider is
$E_0^{eff}\equiv E_0-N\epsilon_b/2$, which we call the effective
ground state energy. This quantity is obtained by numerically solving
the Bethe Ansatz equations (\ref{BAE}) and is plotted as a function of
$1/\gamma$ in Figure \ref{energy}.
Its asymptotic behavior in the BCS \cite{KO} and BEC-limit \cite{LL}
and near the CIR is:
\begin{eqnarray}
\frac{E_0^{eff}}{N\epsilon_F}&\simeq&
\frac{1}{3}\Big[1+\frac{6\gamma}{\pi^2}
-\frac{3\gamma^2}{\pi^4}\log^2(|\gamma|)+\cdots \Big] \text{;
}1/\gamma\rightarrow -\infty \nonumber \\
&\simeq&\frac{1}{12}\Big[1-\frac{1}{\gamma}+\frac{3}{4\gamma^2}+\cdots
\Big] \text{; }1/\gamma\rightarrow 0 \nonumber \\
&\simeq&\frac{\gamma}{\pi^2}\Big[1-\frac{8\sqrt{\gamma}}{3\pi}+\cdots
\Big] \text{; }1/\gamma\rightarrow +\infty
\label{Ener}
\end{eqnarray}
where $\epsilon_F\equiv \pi^2 \hbar^2 n^2/8m$ is the Fermi energy for
the non-interacting gas. We note that on resonance 
$E_0^{eff}=E_0^{eff}(\gamma=0^-)/4$.

We now discuss the low energy elementary excitations. First consider
the density excitations (the ``charge sector'') which are gapless
phonons with dispersion relation $\omega=v_c|k|$, when $|k|\rightarrow
0$. An effective theory is provided by a Luttinger liquid description
with charge velocity $v_c$ and correlation functions described by the
Luttinger parameter $K_c$ \cite{Schulz}. When $\gamma<0$, $K_c^{(F)}$
describes fermionic correlation functions and is obtained from
$K_c^{(F)}=v_F/v_c$ \cite{EL}, where $v_F\equiv \pi\hbar n/2m$ is the
Fermi velocity of the non-interacting gas. When $\gamma>0$,
$K_c^{(B)}$ describes bosonic correlation functions and is given by
$K_c^{(B)}=v_F/2v_c$ \cite{EL}. In order to determine these
parameters, it is therefore enough to extract the sound velocity from
the exact solution.  This is done in the usual way by calculating the
compressibility from the ground state energy \cite{Schulz2}.
\begin{figure}[ptb]
\begin{center} 
\includegraphics[height=4.5cm] {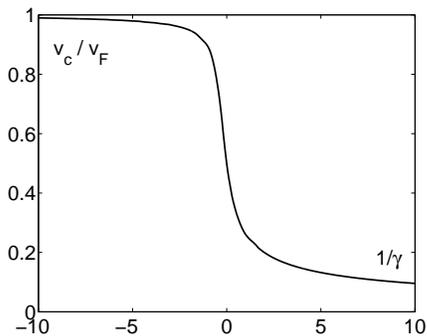}
\end{center} 
\caption{Sound velocity $v_c$ [in units of the Fermi velocity 
for the non-interacting gas $v_F=\pi\hbar n/2m$] as a function of
$1/\gamma$.}
\label{vc} 
\end{figure}
Its asymptotic behavior in the BCS \cite{KO} and BEC-limit \cite{LL}
and near the CIR is:
\begin{eqnarray}
\frac{v_c}{v_F}&\simeq&
1+\frac{\gamma}{\pi^2}+\frac{\gamma^2}{2\pi^4}[\log(|\gamma|)-2]
+\cdots\text{; }1/\gamma\rightarrow -\infty \nonumber \\
&\simeq&\frac{1}{2}\Big[1-\frac{1}{\gamma}+\frac{3}{4\gamma^2}+\frac{3}{4\gamma^3}
+\cdots \Big]\text{; }1/\gamma\rightarrow 0 \nonumber \\
&\simeq&\frac{\sqrt{\gamma}}{\pi}\Big[1-\frac{\sqrt{\gamma}}{2\pi}
+\cdots \Big]\text{; }1/\gamma\rightarrow +\infty
\label{Soundvelocity}
\end{eqnarray}
Figure \ref{vc} shows $v_c$ as a function of $1/\gamma$. Note that
$K_c$ is always larger than one, which implies that the system is
a 1D superfluid, as discussed, e.g., in \cite{KFKPS}.

The low energy spin excitations (the ``spin sector'') are
described by a sine-Gordon model \cite{Schulz2} with coupling
parameter $\beta$. A recent renormalization group analysis of the
sine-Gordon model \cite{Kehrein,spininv} shows that when $\gamma<0$
the system is driven to the strong coupling fixed point
$\beta^2=4\pi$, i.e., the spin Luttinger parameter $K_s$ \cite{Schulz}
is equal to $1/2$. There it becomes equivalent to a non-interacting
massive Thirring model \cite{Coleman}, i.e., to a gas of massive
relativistic fermions. The latter are called massive spinons and may
be interpreted as a quantum soliton of the sine-Gordon model. They
obey the relativistic dispersion relation:
\begin{equation} 
\omega=\sqrt{\left(\Delta/2\hbar\right)^2+(v_s k)^2}
\end{equation}
when $|k|\rightarrow 0$. Therefore, the low energy part of the spin
sector is fully described by the spin velocity $v_s$ and the spin gap
$\Delta$ or, equivalently, by the mass of the spinon
$m_s\equiv\Delta/(2v_s^2)$. The spin gap is also defined as the energy
difference between the singlet ground state and the first triplet
excited state. The spin velocity and the spin gap can be extracted
from the spinon dispersion relation computed from the exact solution
of the GY model. The spin gap has the following limiting
behavior:
\begin{eqnarray}
\frac{\Delta}{\epsilon_F}&\simeq&
\frac{16}{\pi}\sqrt{\frac{|\gamma|}{\pi}}e^{-\pi^2/2|\gamma|} +\cdots
\text{; }1/\gamma\rightarrow -\infty \nonumber \\
&\simeq&\frac{2\gamma^2}{\pi^2}\Big[1-\frac{\pi^2}{4\gamma^2}+\mathcal{O}(\gamma^{-4})
\Big] \text{; }1/\gamma\rightarrow 0^-
\label{Delta}
\end{eqnarray}
\begin{figure}[ptb] 
\begin{center} 
\includegraphics[height=4.5cm] {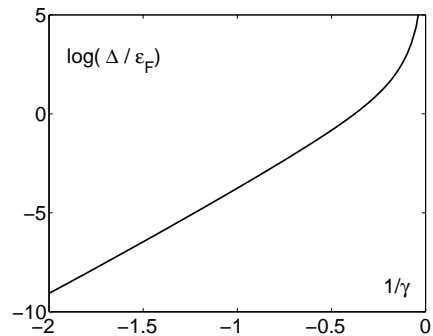}
\end{center} 
\caption{Spin gap $\Delta$ [in units of the Fermi energy for the 
non-interacting gas $\epsilon_F=\pi^2\hbar^2n^2/8m$] as a function of
$1/\gamma$.}
\label{spingap} 
\end{figure}
where the BCS limit was already obtained in \cite{KO}.  When
$\gamma>0$, the spin gap is infinite in our model.  Its behavior as a
function of $1/\gamma$ is plotted in Figure \ref{spingap}.  When
$1/\gamma <-1$, it behaves similarly to the gap computed from the BCS
mean-field theory $\Delta_{BCS}\sim \epsilon_F
\exp{(-\pi^2/2|\gamma|)}$ and the dispersion relation is reminiscent
of that of Bogoliubov quasiparticles, provided $\Delta_{BCS}$ is
identified with $\Delta/2$ \cite{KO}.  Near resonance, the spin gap is
equal to the modulus of the two-body bound state energy
$|\epsilon_b|=2\gamma^2\epsilon_F/\pi^2$. The smooth crossover from
Cooper pairs to molecules occurs when the size of a molecule is of the
order of the average distance between particles $n^{-1}$, which
happens for $\gamma \sim -2$.

\begin{figure}[ptb] 
\begin{center} 
\includegraphics[height=4.5cm] {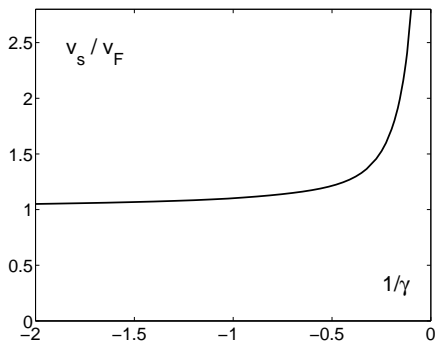}
\end{center} 
\caption{Spin velocity $v_s$ [in units of the Fermi velocity for the non-interacting gas 
$v_F= \pi \hbar n/2m$] as a function of $1/\gamma$.}
\label{spinvelocity} 
\end{figure}

The spin velocity can be computed with the bosonization approach in
the weak coupling limit \cite{Schulz} and from the Bethe Ansatz
equations in the strong coupling limit \cite{KO}. It turns out to be
given by:
\begin{eqnarray}
\frac{v_s}{v_F}&\simeq&1-\frac{\gamma}{\pi^2}+\cdots \text{;
}1/\gamma\rightarrow -\infty \nonumber \\
&\simeq&-\frac{\gamma}{\pi\sqrt{2}}\Big[1-\frac{2}{\gamma}+\cdots
\Big] \text{; }1/\gamma\rightarrow 0^-
\label{vs}
\end{eqnarray}
and is plotted as a function of $1/\gamma$ in Figure
\ref{spinvelocity}. Spin correlation functions can be obtained from
the knowledge of the spin gap $\Delta$ and of the charge Luttinger
parameters $K_c$ and $v_c$, as shown in \cite{OP}.

The above scenario for a BCS-BEC crossover can be realized in an
experiment with ultra-cold gases confined in a quasi-1D geometry,
e.g., $^6$Li in an array of 1D tubes created with optical lattices
\cite{Esslinger} or on an atom chip \cite{Jakob}. In order to reach a
1D regime, the quantum of radial oscillation $\hbar \omega_{\perp}$
has to be much larger than (i) the quantum of axial oscillation $\hbar
\omega_{\parallel}$, (ii) the Fermi energy $\epsilon_F$ and (iii) the
thermal energy $k_B T$. Having a degenerate system imposes that the
latter is also smaller than the Fermi energy. Note that condition (ii)
is required for the validity of the model, as previously
mentioned. Taking, e.g., the following values for radial
$\omega_{\perp}/2\pi\sim 100$~kHz and axial
$\omega_{\parallel}/2\pi\sim 200$~Hz trap frequencies, number of
particles (per tube) $N\sim 100$ \cite{Esslinger} and temperature
$T\sim 50$~nK \cite{GGK} allows to fulfill the previous
requirements. To study the crossover in such a setting, the
dimensionless coupling constant $\gamma$ needs to be tuned through a
Feshbach resonance \cite{GGK}, giving rise to a CIR as discussed
above. The experimental characterization of the different regimes can
be done, for example, by measuring the axial collective modes in the
trap. The ratio of the frequencies of the breathing and dipole modes
\cite{Esslinger} is $2$ in both the BCS and CIR limit \cite{Astra},
and $\sqrt{3}$ in the BEC limit \cite{Menotti}. In addition, extending an 
available RF spectroscopy technique used to measure the molecular binding
energy \cite{RC}, it should be possible to extract information on 
the gap.

In conclusion, we have presented an exactly solvable model for a
BCS-BEC crossover in 1D. The model may be realized experimentally by
using the combination of a standard Feshbach- with a
confinement-induced scattering resonance.  In contrast to the 3D case,
the exact solution allows to make quantitative predictions in the
whole crossover regime. Our results show explicitly that the evolution
from the BCS- to the BEC-limit is continuous, as expected on general
grounds. Of course some of our results are specific for one
dimensional systems, e.g., the pairs are unbreakable on the positive
side of the resonance, which simplifies the description there to a
purely bosonic model. Regarding the behavior of the sound velocity,
however, the situation should be qualitatively similar to the 3D case.
As such, our model provides a novel and experimentally accessible tool
to address one of the long standing basic problems in many-body
physics.

We thank W. Rantner and S. Cerrito for useful discussions and for
carefully reading the manuscript. We also acknowledge discussions with
P. Fedichev, P. Zoller, Z. Hadzibabic and C. Chin.

\emph{Note added}. After this work was completed, we became aware of a
related work of I.V. Tokatly \cite{Tokatly}.


\begin{thebibliography}{99}
\bibitem {sqrt2}Note the $\sqrt{2}$ difference in the definition of
$a_{\perp}$ with \cite{Olshanii,BMO}, which accounts for $A\equiv
-\zeta(1/2,1)/\sqrt{2}\simeq 1.0326$.
\end{thebibliography}
\end{document}